\documentclass{article}
\usepackage{spconf,amsmath,graphicx}
\usepackage[backend=biber,style=ieee,doi=false,isbn=false,url=false,eprint=false]{biblatex}
\usepackage{amssymb}
\usepackage{amsthm}
\usepackage{cancel}
\usepackage{breqn}
\usepackage{multirow}
\usepackage{rotating}
\usepackage{verbatim}
\usepackage{textcomp,gensymb}
\usepackage[caption=false, font=footnotesize]{subfig}

\newcommand{\Expect}{{\rm I\kern-.5em E}}

\DeclareMathAlphabet{\mathitsf}{\encodingdefault}{\sfdefault}{m}{sl}
\DeclareMathAlphabet{\mathitbfsf}{\encodingdefault}{\sfdefault}{bx}{sl}
\usepackage{cool}
\usepackage{flafter}
\title{Decoding 5G-NR Communications via Deep Learning}
\name{Pol Henarejos$^{\star}$, Miguel \'Angel V\'azquez$^{\star}$\thanks{This work has received funding from the ministry of Science, Innovation and Universities under project TERESA-TEC2017-90093-C3-1-R (AEI/FEDER, UE) and from the Catalan Government (2017-SGR-1479).}}

\address{$^{\star}$ Centre Tecnol\`ogic de Telecomunicacions de Catalunya (CTTC), Castelldefels, Spain\\
		Email: \{pol.henarejos,miguel.angel.vazquez\}@cttc.es$^{\star}$}
\begin{document}
%
\maketitle
\begin{abstract}
Upcoming modern communications are based on 5G specifications and aim at providing solutions for novel vertical industries. One of the major changes of the physical layer is the use of Low-Density Parity-Check (LDPC) code for channel coding. Although LDPC codes introduce additional computational complexity compared with the previous generation, where Turbocodes where used, LDPC codes provide a reasonable trade-off in terms of complexity-Bit Error Rate (BER). In parallel to this, Deep Learning algorithms are experiencing a new revolution, specially to image and video processing. In this context, there are some approaches that can be exploited in radio communications. In this paper we propose to use Autoencoding Neural Networks (ANN) jointly with a Deep Neural Network (DNN) to construct Autoencoding Deep Neural Networks (ADNN) for demapping and decoding. The results will unveil that, for a particular BER target, $3$ dB less of Signal to Noise Ratio (SNR) is required, in Additive White Gaussian Noise (AWGN) channels.
\end{abstract}
\section{Introduction}
\label{sec:intro}
Deep Learning science has been studied since many decades. It experienced two technological revolutions in the past and currently, we can affirm that this discipline is empowering the third \cite{Sejnowski2018}. This versatility of Deep Learning algorithms is pushing new applications in many areas of research. 

In particular, Image Processing and Acoustics are studying different types of Deep Neural Networks (DNN) and their benefits to a wide variety of scenarios, such as image classification \cite{Hou2015a}, acoustics classification \cite{Mun2017}, face recognition \cite{Akbulut2017,Zeng2017} and solving regression problems \cite{Lathuiliere2019}. Depending on which neural network is used, different scenarios may arise. For instance, Convolutional Neural Networks (CNN) are suitable for features extraction of hidden patterns \cite{Sainath2013} and Long-Short Term Memory (LSTM) networks are specially indicated for exploiting time series patterns \cite{Karim2018}. All these cases are grouped into the supervised learning area.

In the unsupervised learning area, Autoencoding Neural Networks (ANN) are specially interesting for data compression and reconstruction \cite{Deng2014}. It enables multiple wide applications, such as image compression \cite{Toderici2015} or anomaly detection \cite{Sakurada2014}. The basic idea of ANN is to reduce the dimensionality of the data by consecutive hidden layers and recompose it. The results are particularly interesting not only for the aforementioned cases, but also for denoising \cite{Vincent2008}.

Despite of the majority of the Deep Learning progresses are carried out in the signal processing areas, radio communications can also benefit of the aforementioned approaches. For instance, DNN may be used for interference detection and classification \cite{Henarejos2019}, for theoretical mutual information computation by regression algorithms \cite{Tato2018, Tato2018a}, anomaly and outliers detection \cite{Rajendran2019} or modulation identification \cite{Karra2017,Rajendran2018}, among others.

In this paper, we apply the combination of unsupervised learning of ANN for denoising and supervised learning for regression to signal processing. This approach is introduced by \cite{LinZhou2016,Xie2017} for biomedics and we extend it to signal processing, aiming at increasing the performance of demapping and decoding tasks. In detail, we employ ANN for denoising and highlighting hidden features of radio signals and DNN for regression and prediction of the current wireless standard 5G New Radio (5G-NR) \cite{38212}, which defines the channel coding by using Low-Density Parity-Check (LDPC) codes. The proposed scheme, named Autoencoding Deep Neural Networks (ADNN), replaces the classical approach of symbol demapping and decoding, since both operations are performed by the ADNN. Finally, our results unveil that the combination of ANN and DNN produces lower BER compared with traditional demapping and decoding implementations. 

\section{System Model}
\begin{figure}[!ht]
	\centering
	\includegraphics[width=1\linewidth,clip=true]{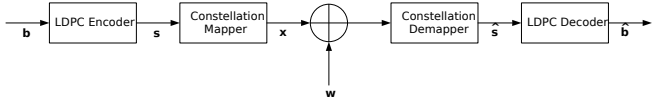}
	\caption{Diagram of system's signal processing blocks.}
	\label{fig:blocks}
\end{figure}

We consider a single-input single-output (SISO) point-to-point wireless system with an Additive White Gaussian Noise (AWGN) channel. The system is composed by a constellation mapper of order $M$, that maps bits to complex baseband symbols, and a channel encoder, composed by a typical LDPC encoder. For a particular time instant $t$, the system model is described as follows
\begin{equation}
y[t]=x[t]+w[t],
\end{equation}
where $y[t]\in\mathbb{C}$ is the received symbol, $x[t]\in\mathbb{C}$ is the encoded symbol and $w[t]\in\mathbb{C}$ is the AWGN. The input vector $x[t]$ is obtained by mapping the bits generated by the LDPC encoder to a constellation known by the receiver. The LDPC encoder has a rate of $(k,n)$, where $n$ bits are produced for every $k$ bits at the input and $k<n$. Fig. \ref{fig:blocks} illustrates the system model.

At the receiver side, the received symbol $y[t]$ is demapped from the constellation and log-likelihood bits are obtained, the \emph{logbits}. The logbits are decoded by the LDPC decoder, whose outputs are the information bits. 

The LDPC channel coding is based on the transmission of blocks of bits. The length of these blocks is denoted by $K$ and, therefore, the length of the output blocks is $L=\frac{n}{k}K$. As explained in the next section, this behaviour of \emph{per-block} basis can be exploited by CNN.

\section{Autoencoding and Deep Neural Networks for Denoising and Decoding}
In this section we introduce the novel approach for denoising and decoding the LDPC encoded blocks, though it can be generalized for other channel codes. ANN can be used for denoising signals and enhancing the SNR. Although it is used in image and speech processing, we propose to apply the fundamental concept of denoising to radio communications systems. After the denoising stage, the DNN decodes the symbols by performing a regression, producing soft decoded bits, the logbits.

The autoencoder is composed by an encoding CNN and a decoding CNN, stacked sequentially. The encoder compresses the data blocks and reduces the dimensionality of the input. The data is passed to the decoder, that increases the dimensions and restores the original dimension. The crucial aspect of this approach is the fact that the autoencoder is trained with known sequences and it is able to recreate the input if it is similar to some of the trained.

CNN are specially indicated for fixed size inputs and they are capable to extract hidden patterns from the data. At the same time, LDPC codes are perfectly suitable for this approach since they have fixed size and they use a parity matrices that introduce predefined patterns, given a particular 5G-NR numerology. Hence, this motivates the use of ANN for denoising radio signals that are encoded with LDPC codes.

The DNN is composed by several hidden layers, whose neurons perform the following operation:
\begin{equation}
\mathbf{y}=\Theta\left(\boldsymbol{\Omega}\mathbf{x}+\mathbf{b}\right),
\label{eq:dense}
\end{equation}
where $\mathbf{y}\in\mathbb{R}^n$ is the output vector, $\boldsymbol{\Omega}\in\mathbb{R}^{n\times k}$ is the weight matrix, $\mathbf{x}\in\mathbb{R}^k$ is the input, $\mathbf{b}\in\mathbb{R}^n$ is the bias and $\Theta\left(\cdot\right)$ is the activation function.

\subsection{Architecture}
Our proposed ANN encoder is composed by several convolutional and decimating layers, placed sequentially:
\begin{enumerate}
	\item Input layer. This layer defines the entry point to the encoder and specifies the size of the input.
	\item Noise layer. This layer introduces Gaussian white noise to the inputs for avoiding overfitting. 
	\item Filter layers. These layers are composed by consecutive convolutional 2D layers and pooling layers. 
\end{enumerate}
Note that filter sizes (i.e., the number of neurons) decrease sequentially by a factor of $2$. For example, the first filter's size is $L$, the following is $L/2$... and successively. 

The ANN's decoder is composed by convolutional layers and interpolating layers, placed sequentially:
\begin{enumerate}
	\item Input layer. This layer defines the entry point of the decoder and the size is the same as the output of the encoder.
	\item Defiltering layers. These layers are the counterpart of filtering layers. 
\end{enumerate}
In contrast to the encoder, filter sizes increase sequentially by a factor of $2$ until the original size $L$ is achieved. 

The DNN is composed by the input layer, $6$ dense layers performing \eqref{eq:dense} and the activation layer.

\begin{figure}[!ht]
	\centering
	\includegraphics[width=1\linewidth,clip=true]{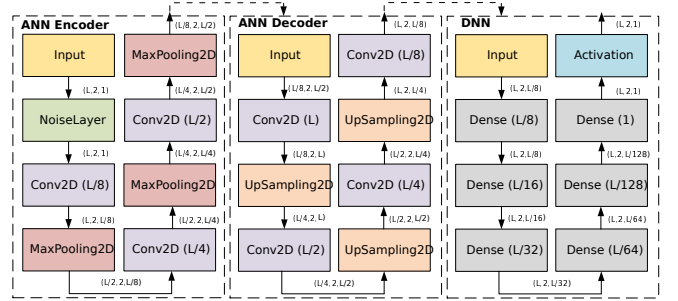}
	\caption{Model of proposed ADNN.}
	\label{fig:model}
\end{figure}
Fig. \ref{fig:model} depicts the proposed architecture of the ADNN for denoising and decoding. The size of data is depicted in brackets between layers. The inputs of the ADNN are the baseband complex symbols (real and imaginary parts) and the outputs are the decoded bits. The labelled names are taken from Keras API of Tensorflow 2.0.

\subsection{Training, Validation and Testing}
One of the most critical aspect of Deep Learning is the training process. During the training, the ADNN is fed with the baseband symbols (namely IQ samples) and the output is compared with bits at transmitting side. All weights and biases are adjusted to produce the output close to the bits at the transmitter. To produce a valid output, first we define the type of loss. In our case, we define two different losses:
\begin{itemize}
	\item Mean Squared Error (MSE): this metric computes the MSE between the output and the input.
	\item BER: computes the BER between the output bits and the input bits at the transmitter.
\end{itemize}

The inputs of the ADNN are the baseband complex symbols, split in two columns (real and imaginary parts). From the neuronal network perspective, all inputs are real numbers, composed in a matrix of shape $(S,2)$.

The processes are repeated sequentially until an acceptable convergence is reached. Each iteration, named \emph{epoch}, refines weights and biases in order to reduce the previous described losses (MSE and BER). To validate the output, a validation data is provided, which is not used during the training but at the end of each epoch for computing the validation losses.

Finally, the last data set is used for testing. This last process uses the input in order to emulate the production environment. At this stage, there is not losses analysis.

\subsection{Noise Layer}
In order to work with noisy scenarios, the training data set has to contain noise. However, since every epoch uses the same inputs, it may derive into overfitting. Using a noisy signal at the input it is equivalent to use the same noise realization at each epoch. To circumvent this, we employ the noise layer in order to generate different realizations for each epoch \cite{Vincent2010}. Hence, the inputs do not contain noise, but it is added by the Noise Layer. This approach decreases the probability to get into overfitting, despite of it is trained for a single SNR value. 

\subsection{Convolution and Pooling Layers}
The Convolution Layer performs the convolution in two dimensions: IQ and time. The convolution process is described by 
\begin{equation}
Y_{i,j}=\sum_{k=0}^{Z-1}\sum_{l=0}^{Z-1}\Omega_{k,l}X_{i+k,j+l},
\end{equation}
where $\textbf{Y}$ is the output, $\textbf{X}$ is the input, $\boldsymbol{\Omega}$ is the kernel and $Z$ is the kernel size. For the sake of simplicity, we use the square convolution. 

After the convolution process, the Pooling Layer takes a value from the output based on a criteria. In our proposal we employ the \emph{max-pooling} criteria, though there are other mechanisms, such as \emph{average-pooling}.

At a given position $k,l$, the max-pooling criteria outputs the maximum value of the input that falls within the kernel. It can be described as
\begin{equation}
Y_{i,j}=\max\left\{X_{i+k,j},\ \forall k\in[0,S)\right\},
\end{equation}
where $S$ is the stride at vertical dimension. Note that we only consider two features, I and Q samples, which are completely independent. Hence, we do not perform max-pooling in the horizontal dimension, only in the vertical dimension.

The Upsampling Layer performs a repetition pattern of the input. It can be expressed as
\begin{equation}
\textbf{Y}_{i,j}=\boldsymbol{1}^T\otimes\textbf{X}_{i,j},
\end{equation}
where $\otimes$ is the Kronecker product, and $\boldsymbol{1}$ is the all-ones vector, whose size is $S$.

\subsection{Activation Layer}
The Activation Layer is the last layer and produces the outputs of the ADNN. There are several activation methods, such as ReLu, sigmoid or hyperbolic tangent. Whilst ReLu and sigmoid are usually used by image processing due to the values of images are positive real numbers, in radio communications the IQ samples are zero-mean, making the distribution centered at $0$. Hence, we employ hyperbolic tangent (namely tanh). 

Since all neurons manage positive and negative real numbers, all neurons use tanh as activation function and the output of each neuron is symmetric with respect to $0$. Hyperbolic tangent has the advantages of symmetry at $0$ but also a continuous derivative, which causes a smooth gradient avoiding discontinuities at the output.

In the next section we implement the proposed ADNN and we simulate a AWGN channel, where the receiver uses the proposed ADNN for denoising and decoding jointly.

\section{Results}
\begin{figure}[!ht]
	\centering
	\includegraphics[width=1\linewidth,clip=true]{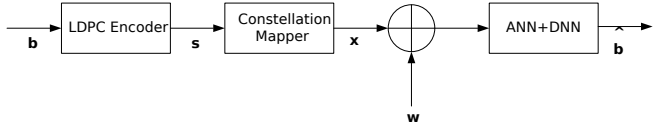}
	\caption{Diagram of the proposed ADNN at the receiver side.}
	\label{fig:ann_blocks}
\end{figure}

In this section we simulate the proposed system. At the transmitter side, we generate $M\times S\times K$ random bits uniformly distributed. The bits are grouped by segments of $K$ length. Each segment is the input of LDPC encoder and its states are reset for every segment. At the output of LDPC encoder, there are $M\times S\times L$ bits. All bits are grouped in $M$ columns of size $S\times L$, corresponding to bits necessary for constellation mapping. We implement the classical approach \cite{Chang2008} depicted in Fig. \ref{fig:blocks}, which is considered the benchmark, and compare with the system proposed in Fig. \ref{fig:ann_blocks}.

The symbols at the output of constellation mapper are the baseband complex symbols, which are passed through an AWGN channel. At the receiver side, we assume perfect synchronization. For the sake of clarity, we do not consider other imperfections (carrier frequency offset, timing misalignments, etc.). 

To test the proposed architecture, we use the framework TensorFlow 2.0. This framework provides an API and a set of layers that can be used for deploying these types of architectures. BER results are plotted by using the MATLAB Python API. In addition to the software, two dedicated GPU NVIDIA RTX 2080 Ti and CUDA 10.1 are used. Thanks to this hardware, all the simulations are boosted by hundreds orders of magnitude, compared to traditional computation in CPUs.

For the simulations, we assume the following hyperparameters: $M=2$, $L=256$, $S=10^5$, training SNR $7$ dB and $100$ epochs. We use two coding rates: $(k,n)=(121,128)$, $K=242$, resulting a coding rate of $R=0.95$, and $(k,n)=(29,32)$, $K=232$, resulting a coding rate $R=0.9$. For the sake of the space, we use two coding rate, though it can be extended to other coding rates. 

The system is tested by a SNR range of $[-6,12]$ dB. Adam optimizer, MSE and BER losses are used for the training process. Note that, even though the ADNN is trained for a single value of SNR ($7$ in our case), the testing process accepts a wide range of SNR at the receiver.

\begin{figure}[!ht]
	\centering
	\subfloat[][Accuracy at each epoch.]{\includegraphics[width=0.48\linewidth,clip=true]{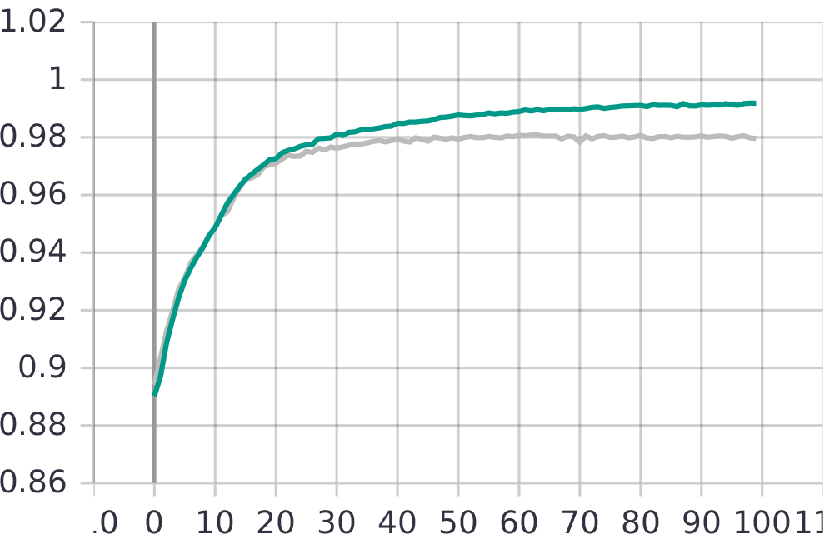}\label{fig:epoch_acc}}\hfill
	\subfloat[][BER at each epoch.]{\includegraphics[width=0.48\linewidth,clip=true]{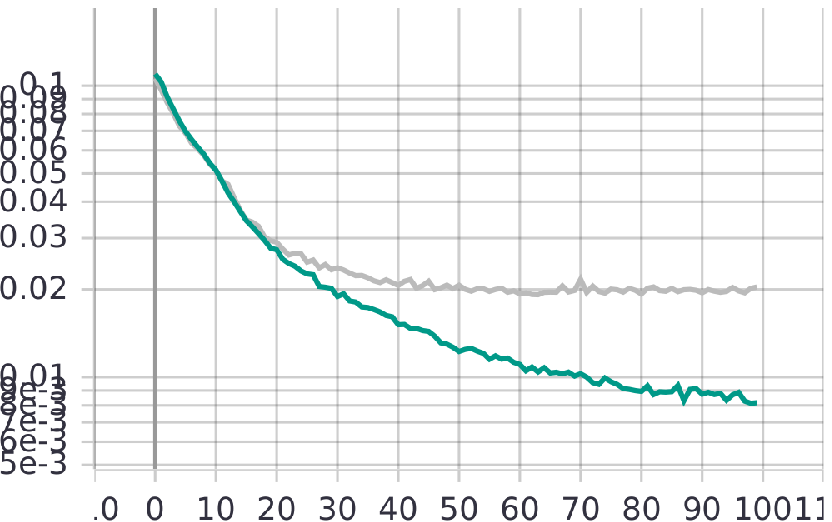}\label{fig:epoch_ber}}\hfill
	\caption{Accuracy and BER of the training process. Green and gray curves correspond to training and validation data sets, respectively.}
	\label{fig:epoch_acc_ber}
\end{figure}
Fig. \ref{fig:epoch_acc} and Fig. \ref{fig:epoch_ber} depict the evolution of the accuracy ($1$ means input completely equal to output) and the BER at each epoch, for training (green) and validation (gray) data sets. At each epoch, the network refines the weights and biases and the convergence is achieved after several decades. Note that, since we are using a noise layer, the accuracy of training cannot reach the maximum. 

\begin{figure}[!ht]
	\centering
	\includegraphics[width=1\linewidth,clip=true]{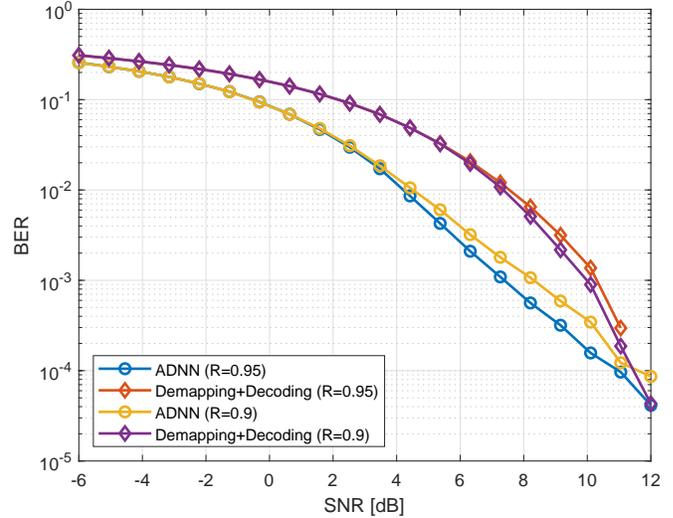}
	\caption{BER of the proposed ADNN compared with classic demapping+decoder \cite{Chang2008}.}
	\label{fig:ber_ae_no}
\end{figure}

Fig. \ref{fig:ber_ae_no} depicts the BER of the proposed architecture (circle marker) compared with the benchmark (diamond marker), an implementation of demapper and decoder by using the Belief Propagation algorithm for AWGN channels \cite{Chang2008}. We can observe that for low and medium SNR the performance of the proposed network is higher compared with the classical approach, which implies that obtains lower BER compared to the benchmark and it specially indicated for low and medium SNR regimes. In detail, at a BER target $10^{-3}$ the proposed scheme requires $3$ dB less of SNR. The immediate consequence is the possibility to use the proposed ADNN for higher coding rates and increase the network throughput. Despite the most time consuming process is the training, once it is finished, the decoding of production is set is faster compared with iterative LDPC decoders.

\section{Conclusions}
In this paper we introduced a novel approach for decoding 5G data frames based on Deep Learning neural networks. Despite of ANN are used for anomaly detection and image compression, we also detailed how the ANN combined with DNN can be applied to future 5G and beyond radio communications. Furthermore, we introduced a meticulous architecture for the implementation of the proposed scheme. Finally, we showed the results of the performance of the proposed system compared with traditional implementations based on constellation demapping and LDPC decoders. Our approach unveiled higher performance and lower BER compared with the classical approach. The results describe that the proposed ADNN can obtain a gain of $3$ dB of SNR. As a future work, other channels can be extended to multipath for exploiting frequency correlations and also to spatial domain through a Multiple-Input Multiple-Output approach.

\section{REFERENCES}
\label{sec:refs}

\printbibliography[heading=none]

@STRING{IEEE_J_NNLS       = "{IEEE} Trans. Neural Netw. Learn. Syst."}

@STRING{IEEE_J_CCN        = "{IEEE} Trans. on Cogn. Commun. Netw."}

@STRING{IEEE_O_ACC        = "{IEEE} Access"}

@Book{Sejnowski2018,
  title     = {The deep learning revolution},
  publisher = {MIT Press},
  year      = {2018},
  author    = {Sejnowski, Terrence J},
}

@Article{Hou2015a,
  author   = {W. {Hou} and X. {Gao} and D. {Tao} and X. {Li}},
  title    = {Blind Image Quality Assessment via Deep Learning},
  journal  = IEEE_J_NNLS,
  year     = {2015},
  volume   = {26},
  number   = {6},
  pages    = {1275--1286},
  month    = jun,
  doi      = {10.1109/TNNLS.2014.2336852},
}

@InProceedings{Mun2017,
  author          = {Seongkyu Mun and Suwon Shon and Wooil Kim and David K. Han and Hanseok Ko},
  title           = {Deep Neural Network based learning and transferring mid-level audio features for acoustic scene classification},
  booktitle       = {IEEE International Conference on Acoustics, Speech and Signal Processing (ICASSP)},
  year            = {2017},
  %%abstract        = {Deep Neural Network (DNN) based transfer learning has been shown to be effective in Visual Object Classification (VOC) for complementing the deficit of target domain training samples by adapting classifiers that have been pre-trained for other large-scaled DataBase (DB). Although there exists an abundance of acoustic data, it can also be said that datasets of specific acoustic scenes are sparse for training Acoustic Scene Classification (ASC) models. By exploiting VOC DNN's ability of learning beyond its pre-trained environments, this paper proposes DNN based transfer learning for ASC. Effectiveness of the proposed method is demonstrated on the database of IEEE DCASE Challenge 2016 Task 1 and home surveillance environment via representative experiments. Its improved performance is verified by comparing it to prominent conventional methods.},
  %date            = {5-9 March 2017},
  doi             = {10.1109/ICASSP.2017.7952265},
  %event%date       = {5-9 March 2017},
  %eventtitleaddon = {New Orleans, LA},
  file            = {:https\://ieeexplore.ieee.org/stamp/stamp.jsp?tp=\&arnumber=7952265:PDF},
  journaltitle    = {2017 IEEE International Conference on Acoustics, Speech and Signal Processing (ICASSP)},
  keywords        = {Acoustics, Training, Neural networks, Conferences, Speech recognition, Surveillance, Convolution, Transfer learning, deep neural network, acoustic scene classification, mid-level feature},
  location        = {New Orleans, LA},
}

@InProceedings{Akbulut2017,
  author          = {Yaman Akbulut and Abdulkadir Sengur and Umit Budak and Sami Ekici},
  title           = {Deep learning based face liveness detection in videos},
  booktitle       = {International Artificial Intelligence and Data Processing Symposium (IDAP)},
  year            = {2017},
  %abstract        = {The human face is an important biometric quantity which can be used to access a user-based system. As human face images can easily be obtained via mobile cameras and social networks, user-based access systems should be robust against spoof face attacks. In other words, a reliable face-based access system can determine both the identity and the liveness of the input face. To this end, various feature-based spoof face detection methods have been proposed. These methods generally apply a series of processes against the input image(s) in order to detect the liveness of the face. In this paper, a deep-learning-based spoof face detection is proposed. Two different deep learning models are used to achieve this, namely local receptive fields (LRF)-ELM and CNN. LRF-ELM is a recently developed model which contains a convolution and a pooling layer before a fully connected layer that makes the model fast. CNN, however, contains a series of convolution and pooling layers. In addition, the CNN model may have more fully connected layers. A series of experiments were conducted on two popular spoof face detection databases, namely NUAA and CASIA. The obtained results were then compared, and the LRF-ELM method yielded better results against both databases.},
  %date            = {16-17 Sept. 2017},
  doi             = {10.1109/IDAP.2017.8090202},
  %event%date       = {16-17 Sept. 2017},
  %eventtitleaddon = {Malatya},
  file            = {:https\://ieeexplore.ieee.org/stamp/stamp.jsp?tp=\&arnumber=8090202:PDF},
  journaltitle    = {2017},
  keywords        = {Face, Databases, Videos, Convolution, Machine learning, Face detection, Heuristic algorithms, Face recognition, face spoof detection, deep learning, CNN, LRF-ELM},
  location        = {Malatya},
}

@InProceedings{Zeng2017,
  author          = {Jinhua Zeng and Jinfeng Zeng and Xiulian Qiu},
  title           = {Deep learning based forensic face verification in videos},
  booktitle       = {International Conference on Progress in Informatics and Computing (PIC)},
  year            = {2017},
  %abstract        = {Deep learning for face identification-verification application has been proven to be fruitful. Human faces constituted the main information for human identification besides gait, body silhouette, etc. Deep learning for forensic face identification could provide quantitative indexes for face similarity measurement between the questioned and the known human faces in cases, which had the advantage of result objectivity without expert experience influences. We studied the deep learning based face representation for forensic verification of human images. Its application strategies and technical limitations were discussed. We proposed a â€œwinner-take-allâ€� strategy in the case of the forensic identification of human images in videos. We expected the theories and techniques for forensic identification of human images in which both qualitative and quantitative analysis methods were included and expert judgment and automatic identification methods were coexisted.},
  %date            = {15-17 Dec. 2017},
  doi             = {10.1109/PIC.2017.8359518},
  %event%date       = {15-17 Dec. 2017},
  %eventtitleaddon = {Nanjing},
  file            = {:https\://ieeexplore.ieee.org/stamp/stamp.jsp?tp=\&arnumber=8359518:PDF},
  journaltitle    = {2017 International Conference on Progress in Informatics and Computing (PIC)},
  keywords        = {Face, Forensics, Machine learning, Face recognition, Videos, Video sequences, Testing, deep learning, forensic identification of human images, face identification, face verification, winner-take-all},
  location        = {Nanjing},
}

@InProceedings{Sainath2013,
  author    = {Sainath, T. N. and Mohamed, A. and Kingsbury, B. and Ramabhadran, B.},
  title     = {Deep convolutional neural networks for {LVCSR}},
  booktitle = {{IEEE} {International} {Conference} on {Acoustics}, {Speech} and {Signal} {Processing}},
  year      = {2013},
  month     = may,
  %abstract  = {Convolutional Neural Networks (CNNs) are an alternative type of neural network that can be used to reduce spectral variations and model spectral correlations which exist in signals. Since speech signals exhibit both of these properties, CNNs are a more effective model for speech compared to Deep Neural Networks (DNNs). In this paper, we explore applying CNNs to large vocabulary speech tasks. First, we determine the appropriate architecture to make CNNs effective compared to DNNs for LVCSR tasks. Specifically, we focus on how many convolutional layers are needed, what is the optimal number of hidden units, what is the best pooling strategy, and the best input feature type for CNNs. We then explore the behavior of neural network features extracted from CNNs on a variety of LVCSR tasks, comparing CNNs to DNNs and GMMs. We find that CNNs offer between a 13-30\% relative improvement over GMMs, and a 4-12\% relative improvement over DNNs, on a 400-hr Broadcast News and 300-hr Switchboard task.},
  doi       = {10.1109/ICASSP.2013.6639347},
  file      = {IEEE Xplore %abstract Record:https\://ieeexplore.ieee.org/%abstract/document/6639347:text/html},
  keywords  = {correlation methods, neural nets, speech recognition, deep convolutional neural networks, large vocabulary continuous speech recognition, spectral variations reduction, spectral correlations model, speech signals, LVCSR tasks, DNN, CNN, convolutional layers, hidden units, pooling strategy, broadcast news, switchboard task, time 400 hr, time 300 hr, Hidden Markov models, Speech, Training, Convolution, Neural networks, Speech recognition, Acoustics, Neural Networks, Speech Recognition},
}

@Article{Karim2018,
  author   = {Karim, F. and Majumdar, S. and Darabi, H. and Chen, S.},
  title    = {{LSTM} {Fully} {Convolutional} {Networks} for {Time} {Series} {Classification}},
  journal  = IEEE_O_ACC,
  year     = {2018},
  volume   = {6},
  pages    = {1662--1669},
  %abstract = {Fully convolutional neural networks (FCNs) have been shown to achieve the state-of-the-art performance on the task of classifying time series sequences. We propose the augmentation of fully convolutional networks with long short term memory recurrent neural network (LSTM RNN) sub-modules for time series classification. Our proposed models significantly enhance the performance of fully convolutional networks with a nominal increase in model size and require minimal preprocessing of the data set. The proposed long short term memory fully convolutional network (LSTM-FCN) achieves the state-of-the-art performance compared with others. We also explore the usage of attention mechanism to improve time series classification with the attention long short term memory fully convolutional network (ALSTM-FCN). The attention mechanism allows one to visualize the decision process of the LSTM cell. Furthermore, we propose refinement as a method to enhance the performance of trained models. An overall analysis of the performance of our model is provided and compared with other techniques.},
  doi      = {10.1109/ACCESS.2017.2779939},
  file     = {IEEE Xplore Full Text PDF:https\://ieeexplore.ieee.org/ielx7/6287639/8274985/08141873.pdf?tp=&arnumber=8141873&isnumber=8274985&ref=:application/pdf;IEEE Xplore %abstract Record:https\://ieeexplore.ieee.org/%abstract/document/8141873:text/html},
  keywords = {learning (artificial intelligence), pattern classification, recurrent neural nets, time series, data set preprocessing, LSTM cell, ALSTM-FCN, LSTM-FCN, FCNs, decision process, fully convolutional neural networks, attention long short term memory fully convolutional network, LSTM RNN, long short term memory recurrent neural network sub-modules, time series classification, Time series analysis, Recurrent neural networks, Feature extraction, Convolution, Computer architecture, Machine learning, Machine learning algorithms, Convolutional neural network, long short term memory recurrent neural network, time series classification},
}

@InProceedings{Henarejos2019,
  author    = {P. {Henarejos} and M. {\'{A}}. {V\'azquez} and A. I. {P\'erez-Neira}},
  title     = {Deep Learning For Experimental Hybrid Terrestrial and Satellite Interference Management},
  booktitle = {IEEE 20th Int. Workshop Signal Processing Advances in Wireless Communications (SPAWC)},
  year      = {2019},
  month     = jul,
  doi       = {10.1109/SPAWC.2019.8815532},
}

@InProceedings{Tato2018,
  author    = {A. {Tato} and C. {Mosquera} and P. {Henarejos} and A. {P\'erez-Neira}},
  title     = {Practical Implementation of Link Adaptation with Dual Polarized Modulation},
  booktitle = {Networks Digital Signal Processing (CSNDSP) 2018 11th Int. Symp. Communication Systems},
  year      = {2018},
  month     = jul,
  doi       = {10.1109/CSNDSP.2018.8471820},
}

@InProceedings{Tato2018a,
  author    = {Tato, Anxo and Henarejos, Pol and Mosquera, Carlos and P\'erez-Neira, Ana},
  title     = {Link {Adaptation} {Algorithms} for {Dual} {Polarization} {Mobile} {Satellite} {Systems}},
  booktitle = {Wireless and {Satellite} {Systems}},
  year      = {2018},
  %abstract  = {The use of dual polarization in mobile satellite systems is very promising as a means for increasing the transmission capacity. In this paper a system which uses simultaneously two orthogonal polarizations in order to communicate with the users is studied. The application of MIMO signal processing techniques along with Adaptive Coding and Modulation in the forward link can provide remarkable throughput gains up to 100\% when compared with the single polarization system. The gateway is allowed to vary the MIMO and Modulation and Coding Schemes for each frame. The selection is done by means of a link adaptation algorithm which uses a tunable margin to achieve a predefined target Frame Error Rate.},
  file      = {Springer Full Text PDF:https\://link.springer.com/content/pdf/10.1007%2F978-3-319-76571-6_6.pdf:application/pdf},
  isbn      = {9783319765716},
  keywords  = {Link adaptation , Adaptive Coding and Modulation , MIMO , Dual-polarization , Satellite communications , Mobile satellite systems },
  language  = {en},
}

@Article{Rajendran2019,
  author   = {S. {Rajendran} and W. {Meert} and V. {Lenders} and S. {Pollin}},
  title    = {Unsupervised Wireless Spectrum Anomaly Detection With Interpretable Features},
  journal  = IEEE_J_CCN,
  year     = {2019},
  volume   = {5},
  number   = {3},
  pages    = {637--647},
  month    = sep,
  doi      = {10.1109/TCCN.2019.2911524},
}

@Article{Rajendran2018,
  author   = {S. {Rajendran} and W. {Meert} and D. {Giustiniano} and V. {Lenders} and S. {Pollin}},
  title    = {Deep Learning Models for Wireless Signal Classification With Distributed Low-Cost Spectrum Sensors},
  journal  = IEEE_J_CCN,
  year     = {2018},
  volume   = {4},
  number   = {3},
  pages    = {433--445},
  month    = sep,
  doi      = {10.1109/TCCN.2018.2835460},
}

@InProceedings{Karra2017,
  author    = {K. {Karra} and S. {Kuzdeba} and J. {Petersen}},
  title     = {Modulation recognition using hierarchical deep neural networks},
  booktitle = {IEEE Int. Symp. Dynamic Spectrum Access Networks (DySPAN)},
  year      = {2017},
  month     = mar,
  doi       = {10.1109/DySPAN.2017.7920746},
}

@Article{Deng2014,
  author    = {Deng, Li},
  title     = {A tutorial survey of architectures, algorithms, and applications for deep learning},
  journal   = {APSIPA Transactions on Signal and Information Processing},
  year      = {2014},
  volume    = {3},
  %abstract  = {In this invited paper, my overview material on the same topic as presented in the plenary overview session of APSIPA-2011 and the tutorial material presented in the same conference [1] are expanded and up%dated to include more recent developments in deep learning. The previous and the up%dated materials cover both theory and applications, and analyze its future directions. The goal of this tutorial survey is to introduce the emerging area of deep learning or hierarchical learning to the APSIPA community. Deep learning refers to a class of machine learning techniques, developed largely since 2006, where many stages of non-linear information processing in hierarchical architectures are exploited for pattern classification and for feature learning. In the more recent literature, it is also connected to representation learning, which involves a hierarchy of features or concepts where higher-level concepts are defined from lower-level ones and where the same lower-level concepts help to define higher-level ones. In this tutorial survey, a brief history of deep learning research is discussed first. Then, a classificatory scheme is developed to analyze and summarize major work reported in the recent deep learning literature. Using this scheme, I provide a taxonomy-oriented survey on the existing deep architectures and algorithms in the literature, and categorize them into three classes: generative, discriminative, and hybrid. Three representative deep architectures – deep autoencoders, deep stacking networks with their generalization to the temporal domain (recurrent networks), and deep neural networks (pretrained with deep belief networks) – one in each of the three classes, are presented in more detail. Next, selected applications of deep learning are reviewed in broad areas of signal and information processing including audio/speech, image/vision, multimodality, language modeling, natural language processing, and information retrieval. Finally, future directions of deep learning are discussed and analyzed.},
  doi       = {10.1017/atsip.2013.9},
  file      = {Full Text PDF:https\://www.cambridge.org/core/services/aop-cambridge-core/content/view/023B6ADF962FA37F8EC684B209E3DFAE/S2048770313000097a.pdf/div-class-title-a-tutorial-survey-of-architectures-algorithms-and-applications-for-deep-learning-div.pdf:application/pdf;Snapshot:https\://www.cambridge.org/core/journals/apsipa-transactions-on-signal-and-information-processing/article/tutorial-survey-of-architectures-algorithms-and-applications-for-deep-learning/023B6ADF962FA37F8EC684B209E3DFAE:text/html},
  keywords  = {Deep learning, Algorithms, Information processing},
  language  = {en},
  publisher = {Cambridge University Press},
  %url       = {https://www.cambridge.org/core/journals/apsipa-transactions-on-signal-and-information-processing/article/tutorial-survey-of-architectures-algorithms-and-applications-for-deep-learning/023B6ADF962FA37F8EC684B209E3DFAE},
  %urldate   = {CURRENT\_TIMESTAMP},
}

@Article{Toderici2015,
  author      = {George Toderici and Sean M. O'Malley and Sung Jin Hwang and Damien Vincent and David Minnen and Shumeet Baluja and Michele Covell and Rahul Sukthankar},
  title       = {Variable Rate Image Compression with Recurrent Neural Networks},
  %date        = {2015-11-19},
  eprint      = {http://arxiv.org/abs/1511.06085v5},
  eprintclass = {cs.CV},
  eprinttype  = {arXiv},
  file        = {:http\://arxiv.org/pdf/1511.06085v5:PDF},
  keywords    = {cs.CV, cs.LG, cs.NE},
}

@InProceedings{Sakurada2014,
  author    = {Sakurada, Mayu and Yairi, Takehisa},
  title     = {Anomaly {Detection} {Using} {Autoencoders} with {Nonlinear} {Dimensionality} {Reduction}},
  booktitle = {{MLSDA} 2Nd {Workshop} on {Machine} {Learning} for {Sensory} {Data} {Analysis}},
  year      = {2014},
  %abstract  = {This paper proposes to use autoencoders with nonlinear dimensionality reduction in the anomaly detection task. The authors apply dimensionality reduction by using an autoencoder onto both artificial data and real data, and compare it with linear PCA and kernel PCA to clarify its property. The artificial data is generated from Lorenz system, and the real data is the spacecrafts' telemetry data. This paper demonstrates that autoencoders are able to detect subtle anomalies which linear PCA fails. Also, autoencoders can increase their accuracy by extending them to denoising autoenconders. Moreover, autoencoders can be useful as nonlinear techniques without complex computation as kernel PCA requires. Finaly, the authors examine the learned features in the hidden layer of autoencoders, and present that autoencoders learn the normal state properly and activate differently with anomalous input.},
  doi       = {10.1145/2689746.2689747},
  file      = {ACM Full Text PDF:https\://dl.acm.org/ft_gateway.cfm?id=2689747&type=pdf:application/pdf},
  isbn      = {9781450331593},
  keywords  = {anomaly detection, auto-assosiative neural network, autoencoder, denoising autoencoder, dimensionality reduction, fault detection, nonlinear, novelty detection, spacecrafts},
  %url       = {http://doi.acm.org/10.1145/2689746.2689747},
  %urldate   = {2019-10-10TZ},
}

@Standard{38212,
  title        = {{TS 38.212; NR; Multiplexing and channel coding}},
  organization = {3GPP},
  year         = {2019},
  booktitle    = {3GPP TS 38.212 version 15.2.0 Release 15},
}

@InProceedings{Chang2008,
  author       = {Chang, Y-M and Casado, Andres I Vila and Chang, M-CF and Wesel, Richard D},
  title        = {Lower-complexity layered belief-propagation decoding of LDPC codes},
  booktitle    = {IEEE International Conference on Communications},
  year         = {2008},
  organization = {IEEE},
}

\end{document}